\documentclass[ammsymb,prl,aps,twocolumn,superscriptaddress,showpacs]{revtex4}

\usepackage{graphicx}

\def \beq{\begin{equation}}
\def \eeq{\end{equation}}
\def \beqar{\begin{eqnarray}}
\def \eeqar{\end{eqnarray}}

\begin{document}

\title{Order and disorder in the Local Evolutionary Minority Game}
\author{E. Burgos}
\email{burgos@cnea.gov.ar, ceva@cnea.gov.ar, rperazzo@itba.edu.ar}
\affiliation{
Departamento de F{\'{\i}}sica, Comisi{\'o}n Nacional de Energ{\'\i }a At{\'o}%
mica, Avenida del Libertador 8250, 1429 Buenos Aires, Argentina}
\author{Horacio Ceva}
\affiliation{
Departamento de F{\'{\i}}sica, Comisi{\'o}n Nacional de Energ{\'\i }a At{\'o}%
mica, Avenida del Libertador 8250, 1429 Buenos Aires, Argentina}
\author{R.P.J. Perazzo}
\affiliation{
Instituto Tecnol\'ogico de Buenos Aires, Departamento de Investigaci\'on y
Desarrollo, Avenida Eduardo Madero 399, Buenos Aires, Argentina}
\date{\today}

\begin{abstract}
We study a modification of the Evolutionary Minority Game (EMG) in which 
agents are placed in the nodes of a regular or a random graph. A 
neighborhood for each agent can thus be defined and a modification of the 
usual relaxation dynamics can be made in which each agent updates her 
decision scheme depending upon the options made in her immediate 
neighborhood. We name this model the Local Evolutionary Minority Game 
(LEMG). We report numerical results for the topologies of a ring, a torus 
and a random graph changing the size of the neighborhood. We focus our 
discussion in a one dimensional system and perform a detailed comparison of 
the results obtained from the random relaxation dynamics of the LEMG and 
from a linear chain of interacting spin-like variables at a finite 
temperature. We provide a physical interpretation of the surprising result 
that in the LEMG a better coordination (a lower frustration) is achieved if 
agents base their actions on local information. We show how the LEMG can be 
regarded as a model that gradually interpolates between a fully ordered, 
antiferromagnetic system and a fully disordered system that can be 
assimilated to a spin glass.
\end{abstract}

\maketitle

\section{Introduction}

There are a great number of situations in which a many agent system self 
organizes by coordinating all the individual actions. A similar situation 
is found in many particle, physical systems. The word ``coordination" that 
is used in a social or economic context is replaced by ``ordering" in the 
study of condensed systems. Examples of self organization are for instance 
the growth of a crystalline structure or a transition leading to some 
specific magnetic phase. 

Interesting situations arise in the case of systems in which the 
optimal configurations for different particles do collide with the each 
other. This is the case for instance in a spin glass in which many spins 
interact through a random interaction. In these cases the system is said to 
display some degree of {\em frustration}. A case of a multi-agent system 
displaying frustration that has been widely considered in the literature is 
the Minority Game (MG) \cite{MGame}. In this model many players have to 
make a binary choice and the winning option is the one made by the 
minority. Frustration may in general arise as a consequence of the 
particular nature of the interactions between the particles as in a spin 
glass, by the boundary conditions imposed by the geometry or the system, or 
the rules to win the game as in the case of the MG. The similarities 
between the MG and spin glass systems has been discussed in great detail by 
Ref. \cite{tano}.  

The macroscopic signature of frustration is that the many agent system can 
not accommodate into a single, optimal state in which the energy is a 
minimum but relaxes instead to one of many, suboptimal configurations that 
correspond to local minima in the energy landscape. These configurations 
display what can be called {\em quenched disorder} i.e. fail to display an 
ordered pattern and the associated fluctuations do not disappear at zero 
temperature.    In a spin glass the orientation of all the spins do not 
reach a unique ordering. In the usual relaxation dynamic that is used to 
model the MG, each player continually modifies her choices and the final 
ordering of all the agents in terms of their respective choices is far from 
unique.

The traditional MG ignores any possible spatial proximity among the 
players. In the present paper we place the agents in a lattice 
\cite{suiza}, \cite{nuestroLMG} thus associating a neighborhood to each 
player. We then let each player to make her decisions depending upon the 
options made by the players of her neighborhood. We next consider a random 
relaxation dynamics in which each player gradually adjusts her probability 
of choosing one given alternative of the binary choice. We call this model 
the Local Evolutionary Minority Game (LEMG). This framework is particularly 
suitable to study the interplay between frustration and size effects as 
well as the emergence of ordered patterns through the relaxation process. 
It also allows a direct comparison with a many spin system in which the 
size of the neighborhood is assimilated to the range of the spin-spin 
interaction.  

In this paper we present and discuss relevant results of numerical 
simulations of the LEMG played in one- and two-dimensional systems. We 
investigate the influence of the value of the size parameter - or what is 
the same, the range of the interaction - in the emergence of ordered, 
optimal configurations. We analyze the case of a one dimensional system 
comparing the LEMG with a model inspired in an antiferromagnetic spin 
system at a finite temperature. At the light of this comparison we show 
that the LEMG can be regarded as a model in which disorder is gradually 
introduced, the control parameter being the range of the interaction. A 
fully disordered system that corresponds to the well known EMG then appears 
to be a particular case in which the range of the interaction is the same 
as the size of the system. For any smaller value of the size parameter the 
system can be assimilated to an antiferromagnet. We use this to extend the 
model to include fluctuations like those produced by a finite temperature. 
This amounts to study the self organization process implied in the LEMG 
including agents that change 1their decision with a finite probability. The 
comparison has the by-product of providing a physical picture to understand 
why a finite range in the interaction among the agents has the effect of 
producing a better coordinated configuration \cite{nuestroLMG} with a lower 
frustration. 

\section {The rules of the game}

\subsection{The EMG}

We first consider the traditional EMG \cite{Johnson}. This involves $N$ 
players that make one binary decision (0 or 1).  Each player has a 
probability $p_i; i=1,2, \cdots , N$ of choosing, say, 0. Each player 
receives one point if her decision places her in the minority and loses 
a point otherwise. We limit ourselves to consider the situation in which 
the amount of the loss is the same as the amount of the gain. The effect 
of relaxing this condition has been considered in \cite{nosotros}.  When 
her account of points falls below a given threshold, she changes $p_i 
\rightarrow p'_i$ with $p'_i \in [p_i- \delta p,p_i+\delta p]$, at 
random, and $\delta p \ll 1$. Reflective boundary conditions are imposed 
at $p_i =0,1$.  All agents are assumed to update the corresponding 
$p_i$'s synchronically.  

It is customary to display the self-organization of the system through the
probability density function $P(p)$ obtained in a statistical ensemble of
systems that are allowed to relax to equilibrium. This function gives the
fraction of the population having a probability between $p$ and $p+ dp$ of
choosing, say, 0. When the probabilistic relaxation is used, the
asymptotic function $P(p)$ is shaped as a U with two symmetric peaks at $p
\simeq 0$ and $p \simeq 1$ thus indicating that the $N$ agents have
segregated into two parties making opposite decisions.  

The relaxation process corresponds to the minimization of an ``energy'' 
function ${\cal E} = \sigma^2/N$ \cite{Thermal} where $\sigma$ is the 
standard deviation of the distribution of groups of agents defined by:

\beq
\sigma^2= \sum _A {\cal P}(A) (A-N/2)^2
\label{costo}
\eeq
where ${\cal P}(A)$ is the probability distribution of groups of $A$
agents that have chosen, say, 0. In \cite{Thermal} it is proven that:
\beq
{\cal E}= \frac{\sigma^2}{N}=  N(<p>-1/2)^2 + (<p>-<p^2>).
\label{costo2}
\eeq

One can thus see that the value of $\sigma^2$ depends upon the properties 
of the above mentioned $P(p)$. At equilibrium $\sigma^2$ is an extensive 
magnitude proportional to $N$.  A minimization of ${\cal E}$ is equivalent 
to find a distribution $P(p)$ with the smallest possible number of losers 
as follows from the fact that $\sigma^2 = <(A-N/2)^2>=<(w-l)^2>/4 =<(N-
2l)^2>/4$ where $w$ ($l$) is the number of winners (losers). If one assumes 
na{\"\i}vely $P(p)=\delta (p-1/2)$ corresponding to a symmetric random walk 
(and also eliminating the term $O(N)$ in Eq.(\ref{costo2}) one gets ${\cal 
E} =1/4$ while $P(p) =$ constant yields ${\cal E} =1/6$.  A better result 
is obtained with the usual random relaxation dynamics for the EMG 
yielding\cite{Thermal} ${\cal E} \simeq 1/8 $. 

Energy and frustration remain linked to each other. For the EMG we can 
define frustration as ${\cal F}= l/N$; which fulfills $0 \leq {\cal F} \leq 
1$. This definition may also be used for any system involving a game with 
multiple players.  The value ${\cal F} = 0$ corresponds to a situation such 
as the ``majority game'' in which a player is a winner if her decision is 
the same as the majority.  This leads to situations that can be assimilated 
to a ferromagnetic phase (all the players (spins) have chosen the same 
option (orientation)). In the EMG there are less winners than losers, and 
therefore $1/2 < {\cal F}_{EMG} \leq 1$.  The lowest possible frustration 
for the EMG is reached when the $N$ (odd) agents are coordinated to produce 
the largest possible minority, i.e.  $(N-1)/2$.  Thus the lowest possible 
frustration for a finite EMG is $ {\cal F}^*_{EMG} =1/2(1+1/N)$.

\subsection {The LEMG}

Within the LEMG  a set of neighborhoods $\{{\cal N}_i\}$ is constructed, 
one for each player. All neighborhoods have the same (odd) number $n$ of 
agents. We consider that the $i-th$ agent belongs to ${\cal N}_i$. We 
define a size parameter by $\zeta =n/N$. We define the set of neighborhoods 
assuming two possible spatial orderings that correspond respectively to a 
one-dimensional chain (1D) or a square two-dimensional (2D) regular grid, 
both with periodic boundary conditions (i.e. corresponding respectively to 
a ring or a torus). When the agents are placed in a regular lattice the set 
$\{{\cal N}_i\}$ can be generated as a sliding (linear or square) window of 
size $n$.

The rules of the LEMG are the same as for the EMG except for the important 
difference that an agent wins or loses points depending whether she is, or 
she is not, in the minority {\em of her own neighborhood}. The $i-$th agent 
pays no attention whatsoever to the agents that do not belong to ${\cal 
N}_i$. The LEMG contains the EMG as a particular case. In fact the LEMG 
with $\zeta=1$ (or $n=N$) is an $N$-fold replica of the usual EMG when 
${\cal N}_i$ coincides with the complete $N$-agent system. In the regular 
orderings in which the neighborhoods are respectively a segment or a square 
with an odd number of agents, the only agent that updates her $p_i$ is 
located at the center of the square or segment. All agents are assumed to 
check their respective neighborhoods and update the corresponding values 
synchronically.

In spite of the fact that each player adjusts her decision according to her 
respective neighborhoods, we will be interested in computing the energy of 
the whole system, i.e. we want to check whether the choices made by each agent 
according to her respective neighborhoods lead the whole ensemble to a more 
efficient coordination scheme. Notice that in principle there are no reasons 
to assume any kind of correlation between local and global optima: an agent 
may be a winner in her neighborhood and a loser when the entire system is 
considered and vice versa.

\section{Relevant features of the LEMG}

\subsection{The density distributions $P(p)$}

\begin{figure}[tbp]
\begin{tabular}{c}
\includegraphics[width=8cm,clip]{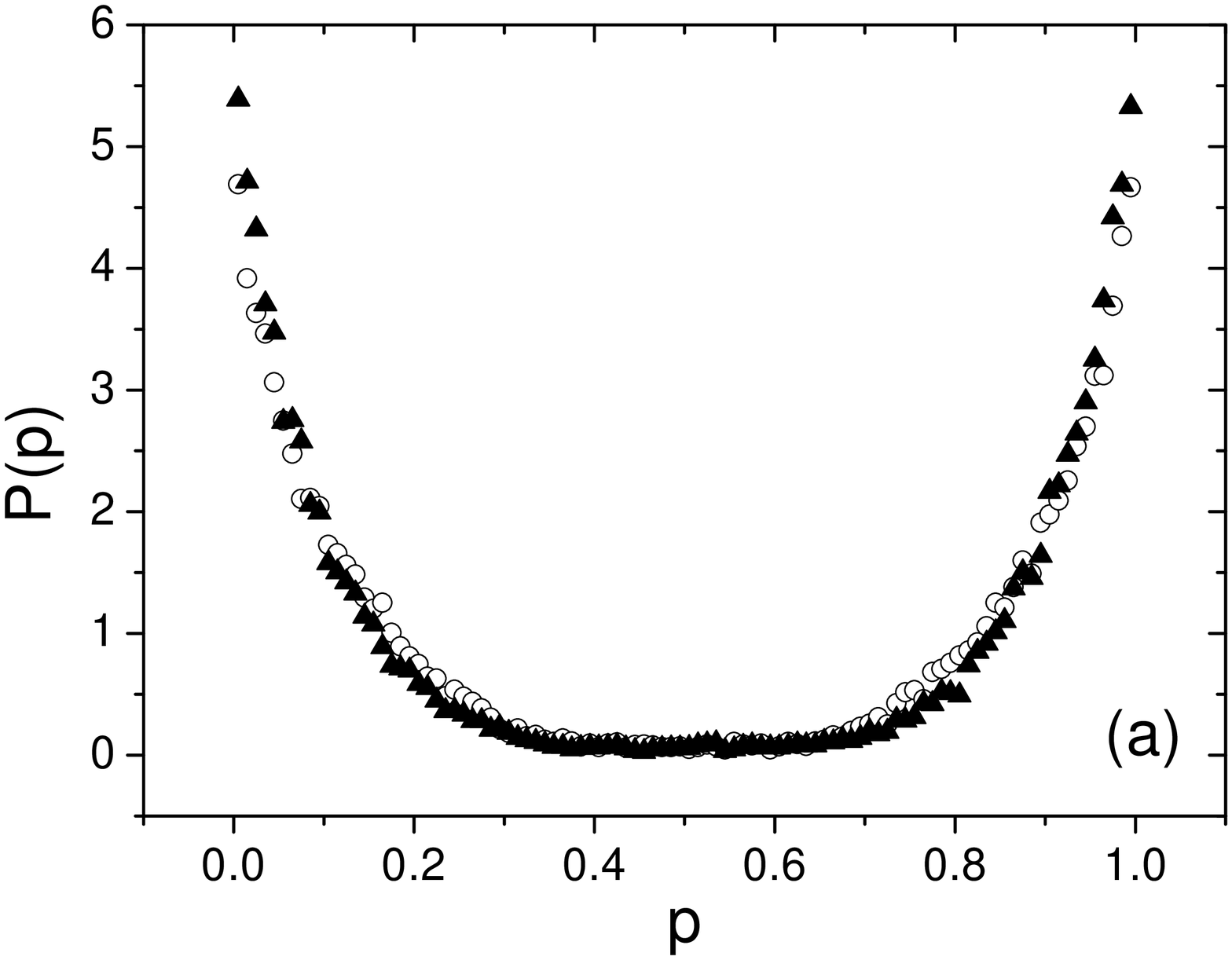}  \cr \includegraphics[width=8cm,
clip]{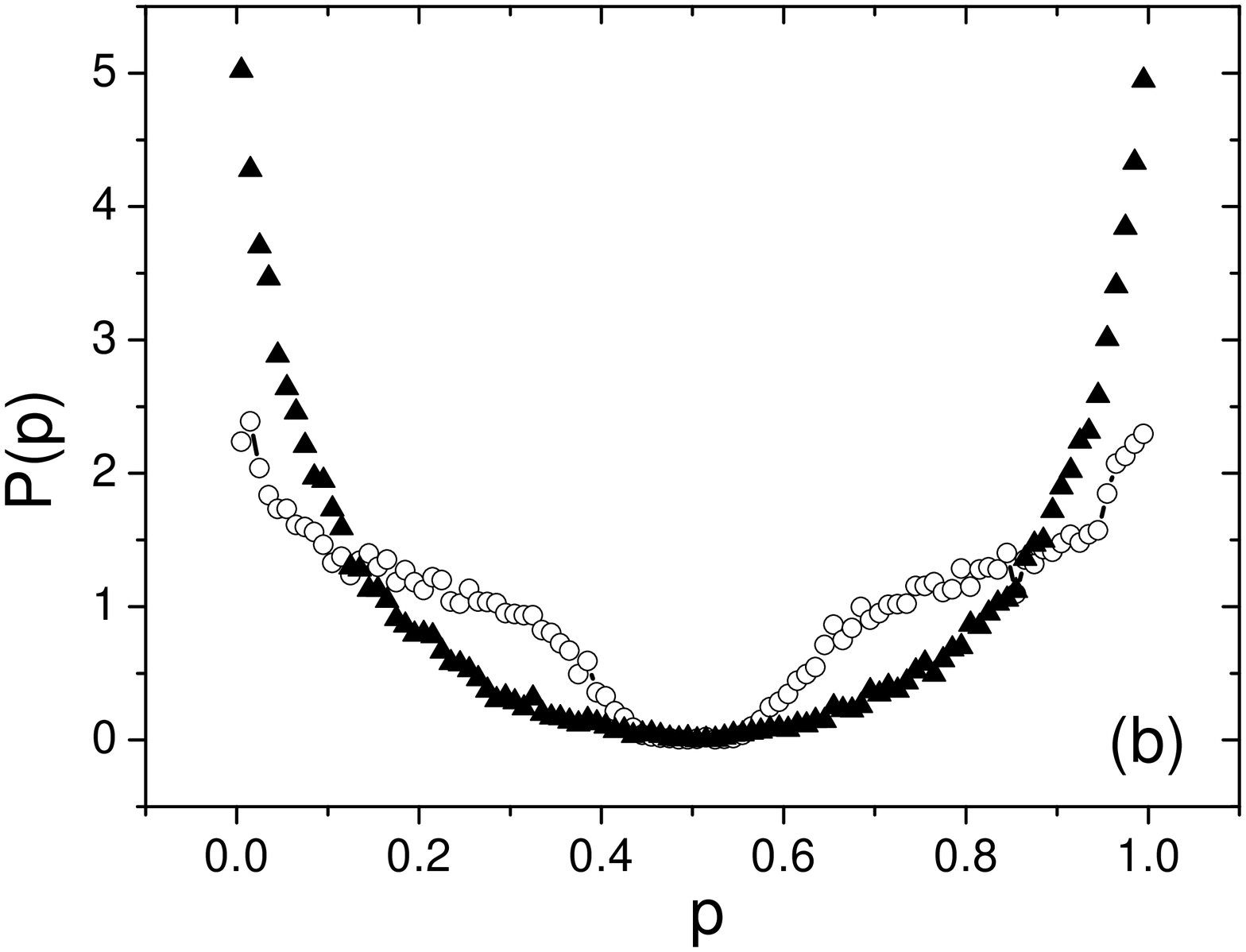} \cr \includegraphics[width=8cm,clip]{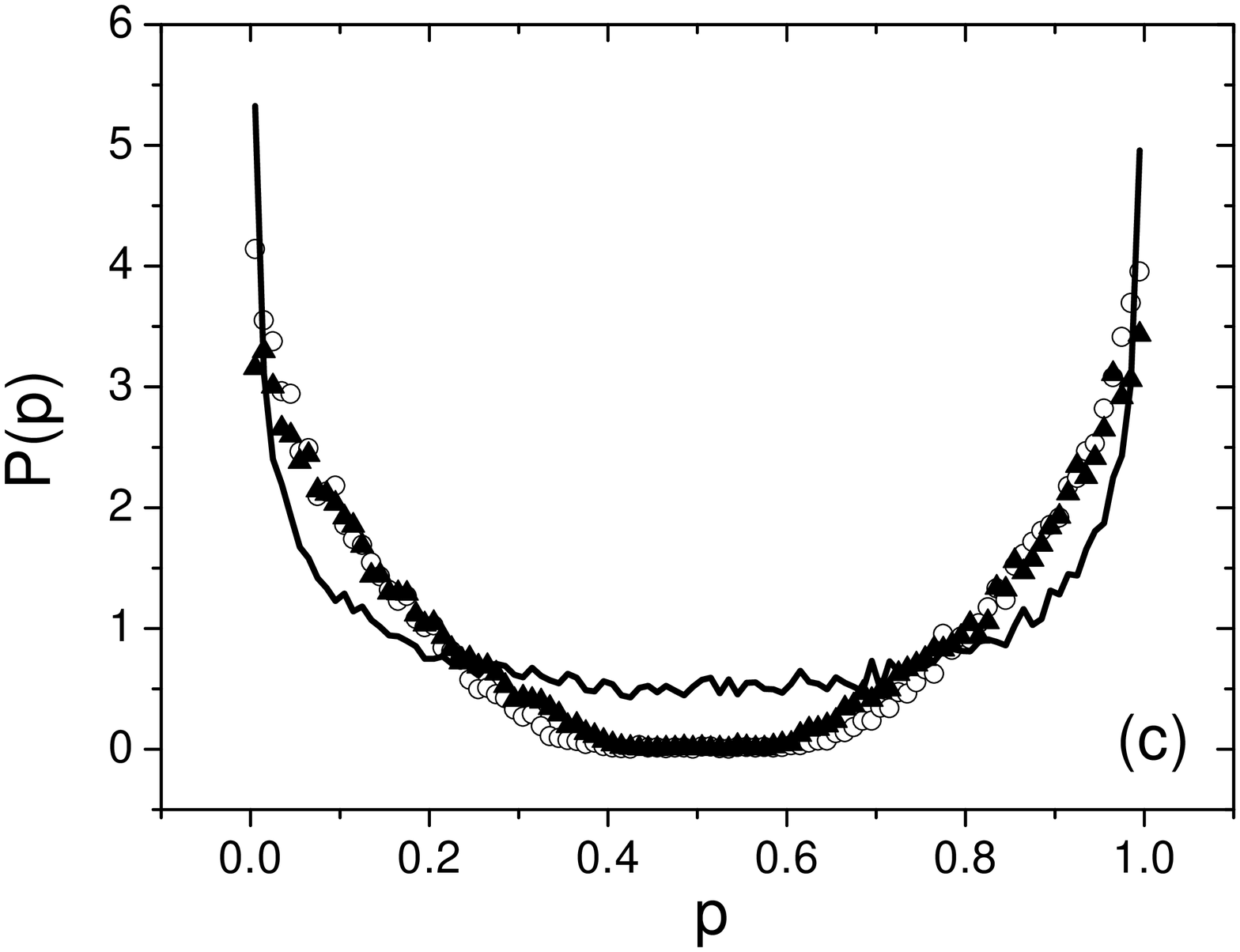} \\
\end{tabular}
\caption{\label {P(p)} Density distribution function for a 1D system with 
$N$=121 agents, and $\zeta$=0.091 (a), 0.587 (b) and 0.950 (c). Open circles 
correspond to ordinary relaxation, while full triangles correspond to an 
annealed relaxation. The full line in (c) corresponds to the EMG, and it is 
shown for comparison.}
\end{figure}

The density distribution functions $P(p)$ that are obtained with the LEMG 
for  1D  systems are shown in Fig.\ref{P(p)}.  The results shown are an 
average over 200 samples, each one involving  5$\times 10^5$ time steps. 
Results for 2D systems are completely similar. The main noticeable 
difference of these functions with those obtained without the neighborhood 
structure is that they drop essentially to zero in an interval that is 
symmetric around $p=1/2$. 

As $\zeta$ grows, the shape of the corresponding $P_{\zeta}(p)$ changes. 
For $\zeta \simeq 1/2$ the distribution has radically changed into the 
peculiar ``two winged'' shape \cite{foot} that can be seen in the figures, 
keeping unchanged the fact that $P_{\zeta}(p \simeq 1/2)=0$. This is 
associated to the rapid stabilization of a noticeable majority of one kind 
of agents in most neighborhoods. 

When this happens the agents are naturally induced to take repeatedly the 
same winning option and keep accumulating points stopping the self-
segregation process that is typical of the MG. The many agent system gets 
``frozen'' (Ref. \cite{Quenching}) in a configuration that is far from the 
``better'' {\em local} optimum. A suboptimal configuration of this kind can 
be improved resorting to an annealing procedure (see Ref.\cite{Quenching}) 
that amounts to periodically remove the points accumulated by all the 
players resetting their accounts to 0. In Fig.\ref{P(p)} the annealing 
procedure is applied every 500 time steps during the first 4$\times 10^5$ 
updatings. It can be seen that the annealing procedure is irrelevant for 
values of $\zeta$ of the order of $\simeq 0$ and $\simeq 1$. The 
distribution functions obtained after the application of such annealing 
procedure regain the well known U- shape that displays the self-segregation 
of the ensemble of agents into two subpopulations with $p_i \simeq 0$ and 
$p_i \simeq 1$ as a consequence of the relaxation dynamics. 

The distributions obtained within the LEMG with or without annealing 
fail to reveal the neighborhood structure of the LEMG. They are unrelated 
to the underlying neighborhood structure, do not signal the emergence of a 
long range order and therefore fail to provide any picture of the changes in 
the emergent ordering of the system as the size parameter changes. 

\subsection{The emergence of a long range order}.
\label{orden} 

\begin{figure}[tbp]
\begin{tabular}{cc}
\includegraphics[width=4cm,clip]{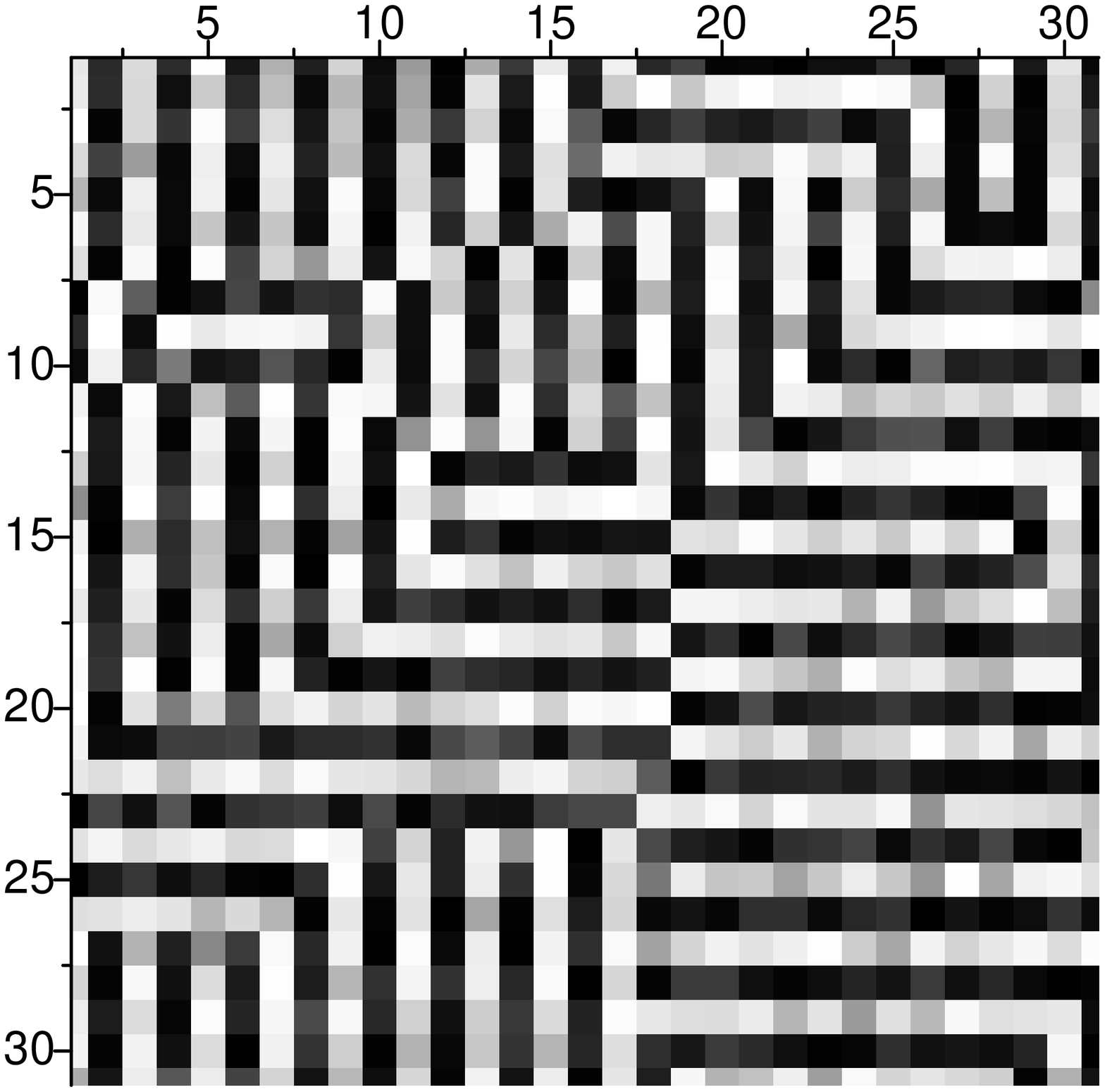} &  \includegraphics[width=4cm,clip]
{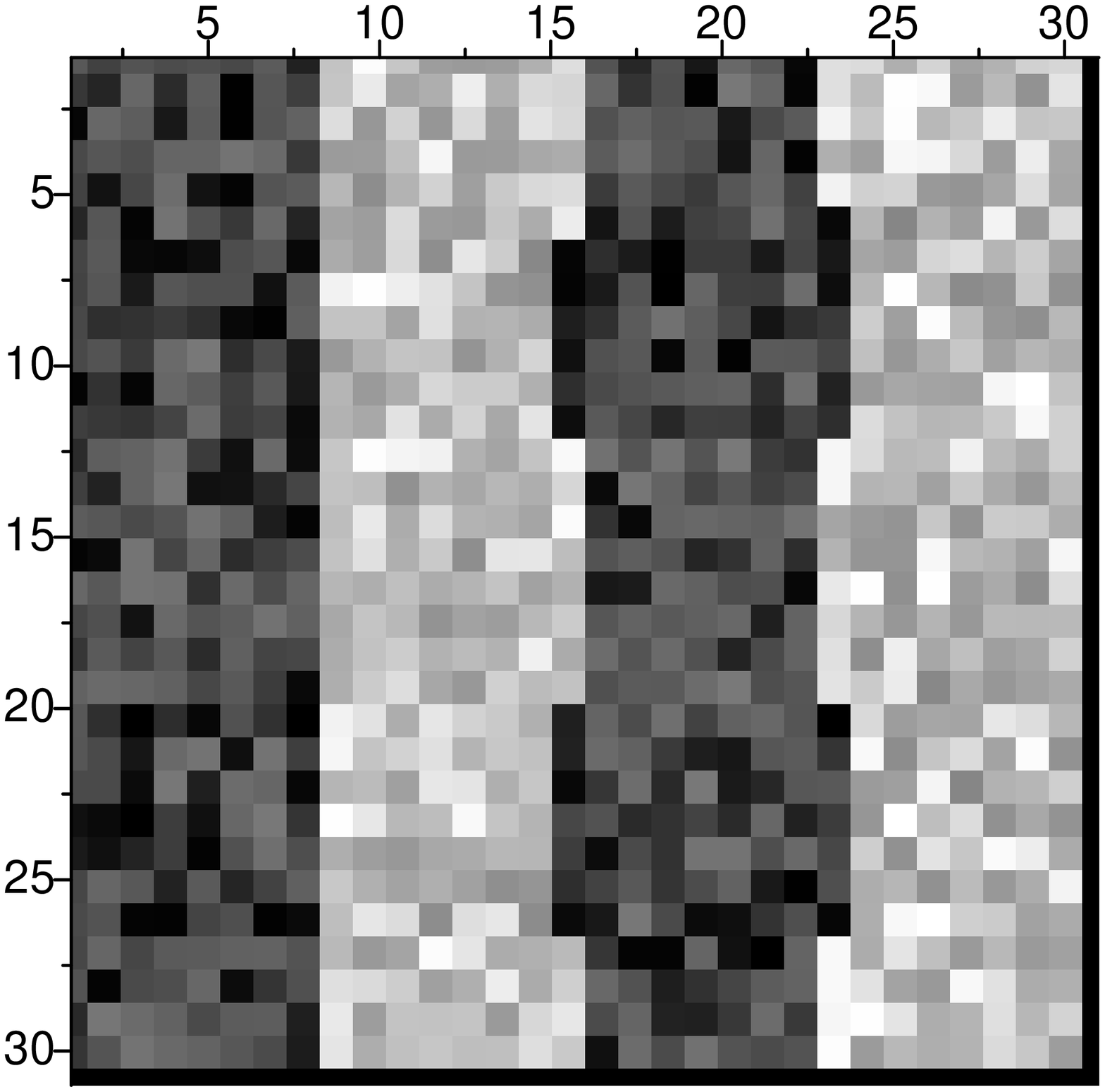} \\
\end{tabular}
\vspace{-1.5cm}
\caption{Results of the relaxation dynamics for a 2D system of 961 agents
located in a grid of 31$\times$31 sites after $10^6$ time steps. The values of 
the $p_i$ are shown in shades of gray. Left panel: neighbourhood of 3$\times$3 
sites. Rigth panel: neigbourhood of 29$\times$29 sites. Domains are clearly 
noticeable in the left panel, while on the rigth panel the whole system is in a 
single domain}  
\label{franjas y escaleras}
\end{figure}

In Fig.\ref{franjas y escaleras} we present the results obtained for the 
case in which the agents are placed in a two dimensional regular lattice. 
We assume $N= K^2$ ($K$ odd) and $n=k^2$ ($k$ odd). We show the values of 
the $p_i$'s at the end of the relaxation process by shades of gray. In all 
cases a pattern arises in which a long range order prevails that depends 
upon the type and the size of the neighborhood that has been chosen. If 
the neighborhood is assumed to contain only the four players that are N, 
S, E and W of the player updating her $p_i$, a regular checkered pattern 
emerges that may also alternate with a pattern of ``stairs''. If a complete 
neighborhood of eight neighbors is chosen the associated ordered pattern is 
one of vertical or horizontal stripes. For some values of $k$ ($k \ll K$) 
the stripes have indented edges. 

These regular patterns are in turn grouped into domains with rather sharp 
boundaries. These ``dislocations'' are a signature of conflicting boundary 
conditions (if $K$ is odd the square with periodic boundary conditions can 
not accommodate an even number of stripes). Most of the agents that are 
``local losers'' - and therefore continue to update their $p_i$'s - lie 
precisely on these borders giving therefore rise to a dynamics in which the 
domain borders slowly move. 

As the size parameter grows the whole grid becomes a single domain with 
wider stripes and the domain structure naturally disappears. Beyond a 
critical value of $\zeta $ the whole grid merges into a single domain. As 
$\zeta $ becomes even larger the edges of the stripes have noticeable 
indentations. Finally these dominate the whole picture and when $\zeta 
\simeq 1$ all traces of a regular ordering completely disappear. 

\begin{figure}[tbp]
\begin{tabular}{c}
\includegraphics[width=8cm,clip]{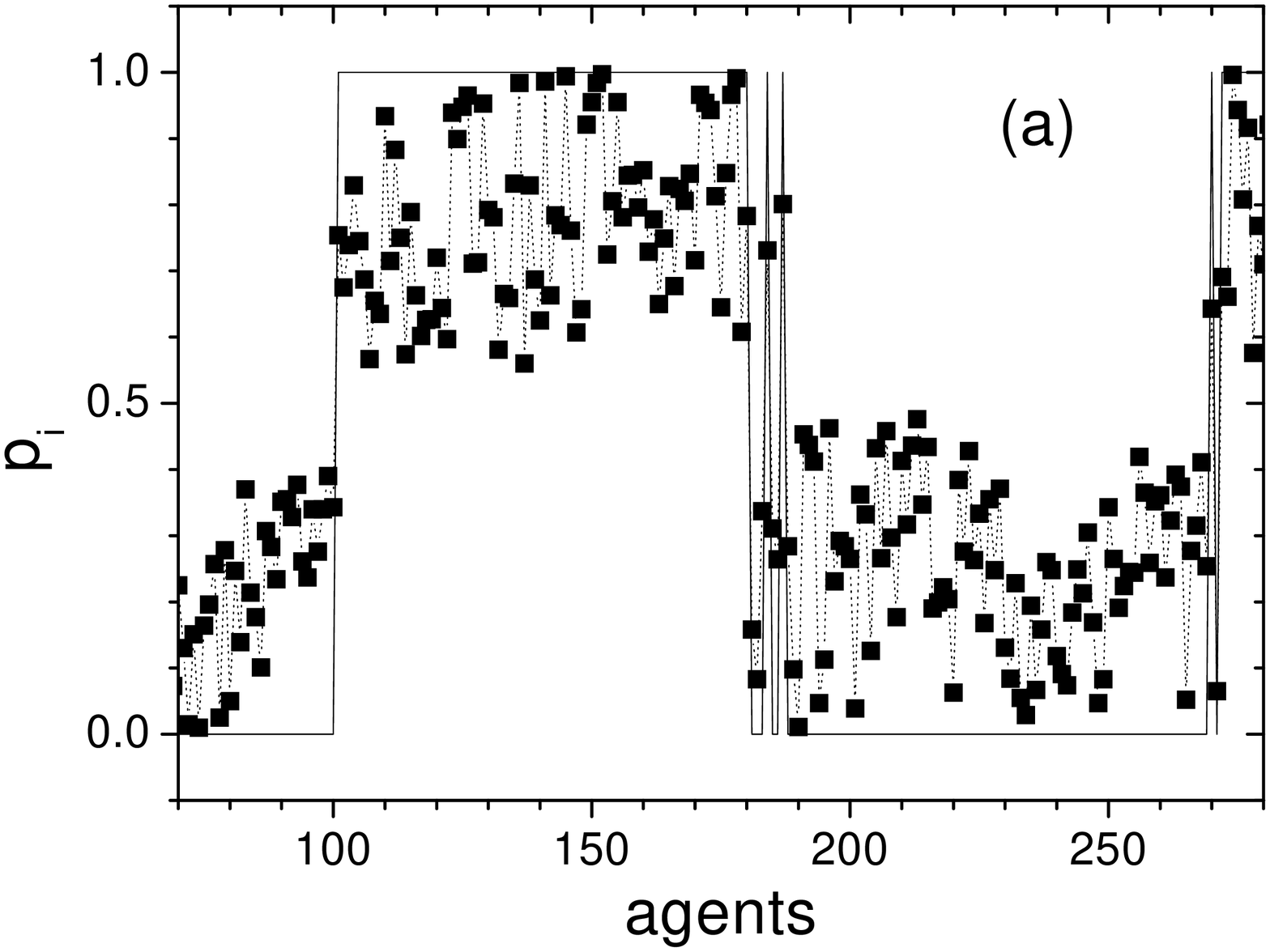} \cr \includegraphics[width=8cm,
clip]{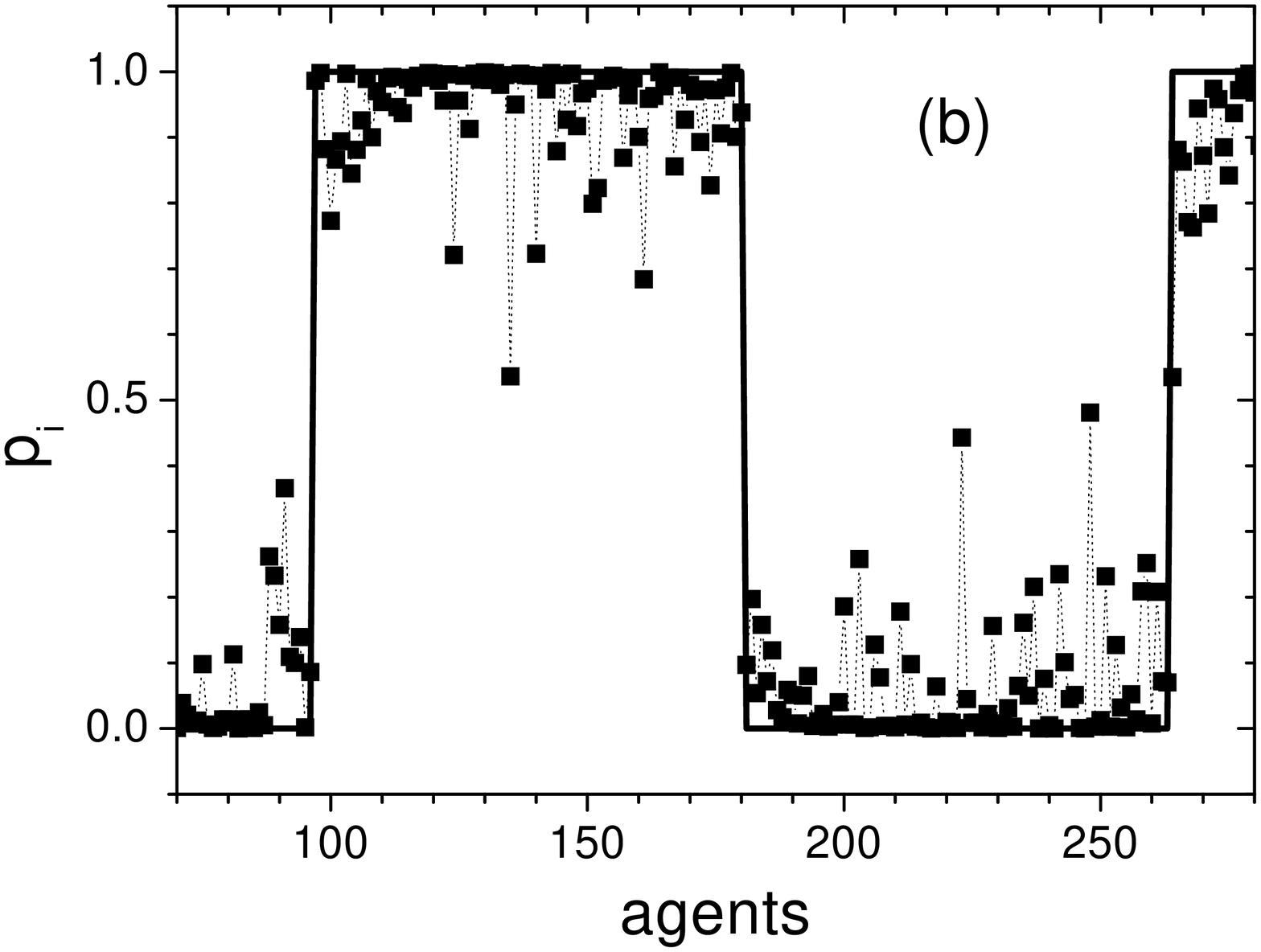} \cr \includegraphics[width=8cm,clip]{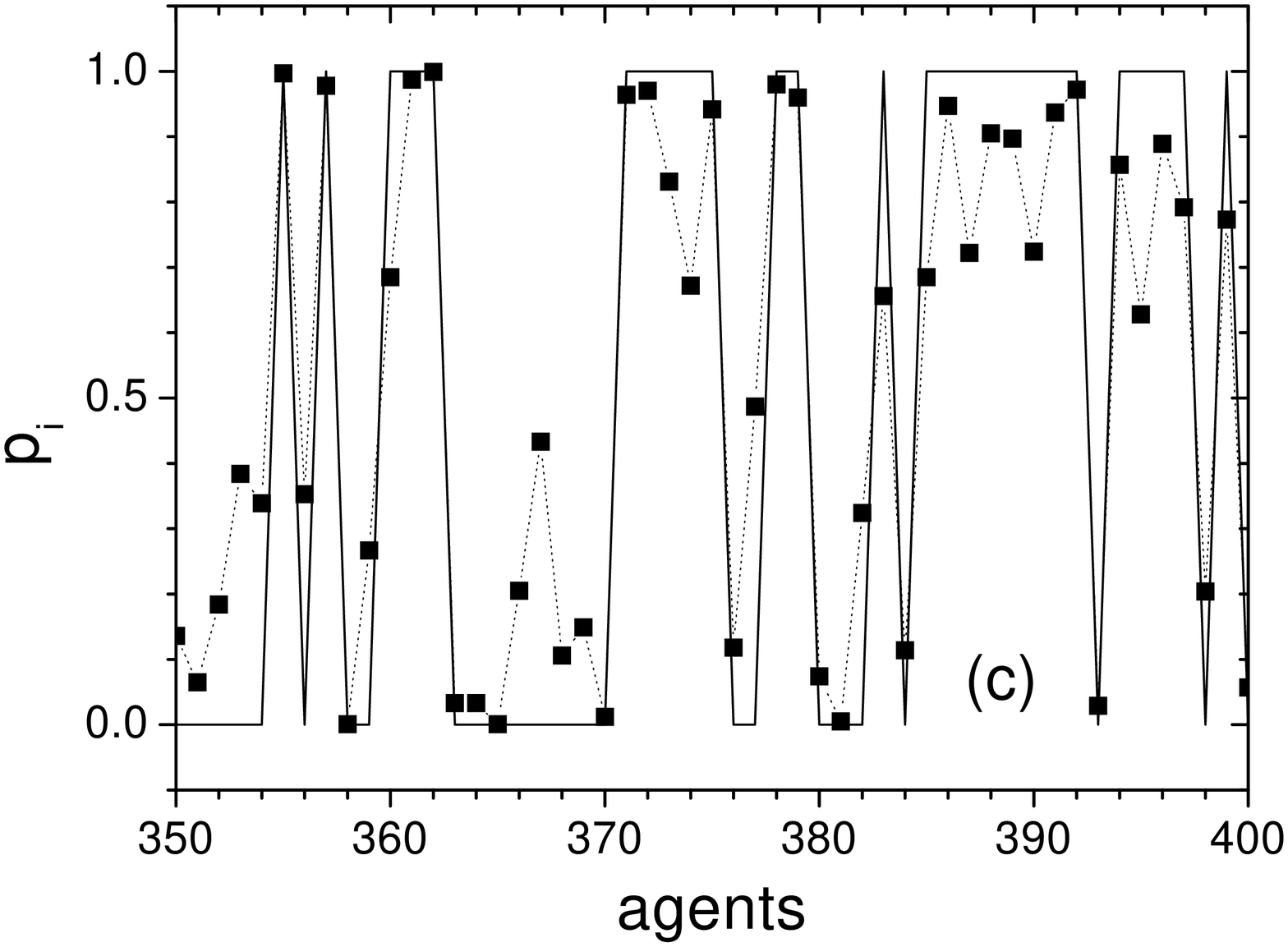} \\
\end{tabular}
\caption{Results of the relaxation dynamics for a 1D system of 501 agents, 
after 5$\times 10^5$ time steps. We plot the individual probabilities $p_i$ 
of the agents along the line (full squares). For clarity we show only a small, 
representative portion of the system. The doted lines are a guide to the eye. 
The full line is the result of rounding the value of $p_i$ to zero or one. 
Panels (a) and (b) correspond to a neighborhood of 251 agents, while panel (c) 
corresponds to one of 501 agents, \emph{i.e.} the EMG. Panel (b) is the 
annealed version of panel (a). Notice the periodicity displayed in (a) and (b), 
while the EMG result shows no long range order}
\label{cadena 1D}
\end{figure}

Domains are a peculiar feature of 2D and higher dimensional system. They do 
not appear in 1D chains. In Fig \ref{cadena 1D} we show some typical 
results for a 1D system. In 1D systems one can not find a domain structure 
in which portions of the chain in which one periodicity prevails are side 
by side with another portion in which the prevailing period is different. 
For larger values of $\zeta$ strings of consecutive players with nearly 
opposite, extreme values of $p_i$, all having lengths that are very close 
to a given length, alternate with each other randomly. In addition, this 
privileged length is close to $n/3$. 

It is interesting to note that the annealing procedure mentioned in the 
preceding section does not change in any way the long range ordering of the 
system. As can be seen in Fig.\ref{cadena 1D} an improvement of the optimum 
is achieved only by forcing the agents that already have $p_i > 1/2$ or 
$p_i < 1/2$ to have $p_i$'s that are closer respectively either to 1 or to 
0. As $\zeta$ approaches 1 the long range ordering is seen to brake down: 
strings of consecutive players with opposite values of $p$ change their 
lengths randomly.

\subsection{Energy minima} 

\begin{figure}[tbp]
\includegraphics[width=9.5cm,clip]{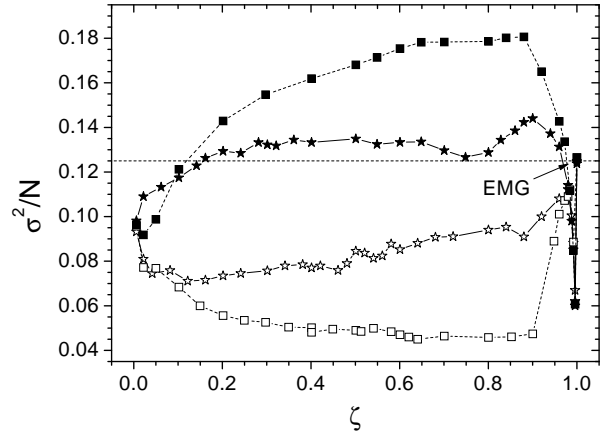}  
\caption{$\sigma^2/N$ as a function of the size parameter $\zeta$ for 
different topologies of the system of $N=501$ agents. Lines are drawn to
guide the eye. Stars and squares  correspond to 1D and random 
graphs, respectively. Empty and filled symbols correspond  to 
results obtained with and without annealing (see the text).}
\label{energia}
\end{figure}

We now consider the total energy ${\cal E}= \sigma^2/N$. In 
Fig.\ref{energia} we show results for  ${\cal E}$ as a function of 
$\zeta$ obtained in several numerical experiments. The value for $\zeta 
=1$, labeled ${\cal E}_{EMG}$, is the one corresponding to the EMG. We show 
the results of the topologies of a ring and a random system.  Data was 
obtained from 20 independent samples, each one of $5 \times 10^5$ time 
steps, by averaging over the last 2000 time steps of all the samples. The 
annealing procedure was applied as explained before. In a preceding paper 
\cite{nuestroLMG} we have pointed out the fact that the total energy and 
hence the degree of coordination of the $N$ agents is always better for the 
LEMG than for the EMG provided that $\zeta\simeq 1$ or $\zeta \simeq 0$. In 
addition if the annealing procedure is followed it turns out that ${\cal 
E}_{\zeta} < {\cal E}_{EMG}$ regardless the value of $\zeta$.

Quenched suboptimal configurations can be improved through annealing, 
making it possible to reach a``better'' local configuration. The 
composition of the ``good'' local optima that are obtained in this way 
always yields values of ${\cal E}_{\zeta}$ that are significantly lower 
that ${\cal E}_{EMG}$. A typical value of ${\cal E}_{\zeta} \simeq 1/16$ is 
obtained in this way that is half the value $\sigma^2/N \simeq 1/8$ 
obtained for the stochastic relaxation dynamic. This remarkable results 
implies that when agents adjust their behavior taking into consideration 
only the agents of their immediate neighborhood (i.e. having a limited 
information about the system) they reach a better coordination than when 
each player considers the behavior of the total ensemble of players (i.e. 
has a better information about the system as a whole).

\section{The 1D case}

The salient features of the LEMG that we have reported in the preceding 
section can be summarized as follows. The stationary configuration that is 
reached at the end of the relaxation process, displays a long range order 
that is related to the size of the neighborhood $n$. A better understanding 
is desirable about that relationship as well as the collapse of the 
internal order as $\zeta \rightarrow 1$. Finally it is necessary to 
supplement the distribution function $P(p)$ with a more reliable indication 
of that internal order. The density function $P(p)$ clearly displays the 
self-segregation of the population but is not sensitive to the structure of 
neighborhoods. A similar statement can also be made about the energy ${\cal 
E}$ introduced in Eq.(\ref{costo2}). In spite of the fact that this 
function turns out to be sensitive to the neighborhood structure, ${\cal 
E}$ is not an explicit function of it. In order to clarify these points we 
will focus our discussions on the 1D case and we will explore many 
similarities between the LEMG and a system of interacting spins inspired in 
an antiferromagnetic Ising model. This can be tailored to be more sensitive 
to the structure of neighborhoods and also allows a thermodynamic treatment 
through the introduction of a temperature and a free energy.

\subsection{The periodic patterns}

Let us consider an infinite linear chain of agents and a neighborhood with 
$n=3$. Let $R_i$ be the probability that the $i$-th agent belongs to the 
minority of ${\cal N}_i$. We can thus write, for $i=-\infty,\cdots,\infty$:

\beq
R_i= (1-p_{i-1})p_i(1-p_{i+1})+p_{i-1}(1-p_i)p_{i+1} 
\label{probentorno}
\eeq

The probability that all agents are winners is

\beq
R=\prod_i R_i.
\eeq

Obviously $R=1$ if and only if $R_i=1,\forall i$ . This is possible only in 
the case in which $p_ i=1$ and $p_{i\pm 1}=0$ (the opposite case in which 
$p_ i=0$ and $p_{i\pm 1}=1$ is completely equivalent). This situation 
corresponds to an ordered pattern in which 0's and the 1's alternate with 
each other. This is a case in which the alternating strings are of unit 
length, that in turn corresponds to $n/3$.  The case of an infinite linear 
chain is therefore not frustrated but any finite linear chain with periodic 
boundary conditions and with an odd number of agents is frustrated.   

It is possible to use Eq.(\ref{probentorno}) to construct a (discrete time) 
relaxation dynamics to adjust the $p_i$'s. If we assume that 
$p_i(t+1)=p_i(t)+\delta p_i$ we can write  

\beq
R_i(t+1) \simeq R_i+\frac{\partial R_i}{\partial p_i}\delta p_i
\eeq

we can therefore insure that $R_i(t+1)-R_i(t)\geq 0$ by choosing

\beq 
\delta p_i=\eta \frac{\partial R_i}{\partial p_i}=\eta (1- p_{i-1} -p_{i+1})
 \ \ \ ;\ \ \eta>0 
\label{dinamica}            
\eeq

A possible stationary solution ($\delta p_i =0 \ \forall i$) of 
Eq.(\ref{dinamica}) is $p_i=1/2\ ;  \forall i$. One can immediately 
recognize that this solution is unstable because any random, small 
perturbation of any $p_i$ leads to a dynamics in which $\delta p_i \not= 0; 
\forall i$. There are other stationary patterns such as the saw tooth 
profile  that repeats the pattern $p_i = 1/2-\epsilon$, $p_{i+1}= 0$, 
$p_{i+2} = 1/2+\epsilon$, $p_{i+3}= 0$, $p_{i+4} = 1/2-\epsilon$, etc. One 
can check that these solutions are also unstable. This dynamics stabilizes 
a pattern of 0's and 1's that alternate with each other.  In fact if 
$p_{i+1}$ and $p_{i-1}$ are both greater (smaller) than 1/2, then $\delta 
p_i < 0 (>0)$ thus forcing $p_i <1/2 (>1/2)$. In addition this relaxation 
dynamics leads to distributions $P(p)$ that vanish at $p=1/2$ as we have 
seen in the numerical experiments reported in the preceding section.  

Larger (odd) neighborhoods give rise to periodic patterns of alternating 
strings of 0's and 1's. It is simple to check by inspection that for a 
neighborhood of length $n=5$ the length of the strings must be 2. However 
for $n=7$, alternating strings of various lengths (1 ,2 and 3) are 
admissible. The coexistence of several periodic solutions for these rather 
small values of $n$ is the responsible for the indented edges of the 
stripes that appear in 2D systems with a neighborhood with $n=49$ (i.e. a 
square neighborhood of 7 sites in each direction). It is possible to verify 
by direct inspection that a concatenation of strings of length $n/3$ tends 
to insure that players always belong to their local minorities. 

For large values of $n$, the relaxation dynamics leads to asymptotic 
stationary states in which shorter string lengths do not survive. One can 
understand this effect on the grounds that a concatenation of short strings 
is highly unstable against fluctuations in their lengths. When longer strings 
are involved, a misadjustment by which there are neighboring strings with a 
length differing in one unit, introduces few losers. A misadjustment of the 
same kind when short strings are involved causes a larger sequence of losers.

\subsection{A spin model for the LEMG}
\label{spinmodel} 

The LEMG on a linear chain gives rise to a specific distribution of the 
lengths of strings of players that make similar decisions. This feature can 
fruitfully be studied by mapping the LEMG into a spin model related 
to an Ising antiferromagnetic model. We consider a linear chain of spins 
${s_i},i=1,2,\cdots ,N$ with $s_i \in \{-1,+1\}$ and a system of 
neighborhoods of size $n$ as considered above. In order to compare this 
model to the LEMG, each spin is assimilated to an agent and its twofold 
orientation is taken to correspond to the binary option for each agent. 

The Hamiltonian function that is a minimum when each agent belongs to 
the minority of her neighborhood is:

\beq
H=\frac{1}{4}\sum_i^N s_i\quad\Bigl[\frac{1}{n-1}\sum_{j\in{\cal N}_i;j
\neq i}s_j \Bigr]\quad \equiv \frac{1}{4}\sum_i^N s_i Q_i
\label{eneloc}
\eeq

The factor $(n-1)^{-1}$ is introduced to have a unified scale of energies 
independent of the value of the size parameter. The remaining factor 1/4 
is instead introduced in order to render possible the correspondence of 
$H$ with ${\cal E}$ in the limit $\zeta \rightarrow 1$. When $n=N$, $H$ 
can be written as:
\beq
H=\frac{1}{4(N-1)}\Big[ (\sum_i^N s_i)^2-\sum_i^Ns_i^2 \Big].
\eeq
This expression can be cast into the form of Eq.(\ref{costo2}) by assuming 
that $N \gg 1$ and that the discrete spin variables $s_i$ and the continuous 
probability variables $p_i$  can be associated with each other through 
\cite{foot2} $s_i=2p_i-1$ .  Notice that the energy given by Eq.(\ref{eneloc}) 
provides an extension of ${\cal E}$ with an explicit dependence upon the 
neighborhood structure. This causes that a low value of $H$ given by Eq.
(\ref{eneloc}), for some given value of $n$ corresponds also to a low value 
of ${\cal E}$ but the reciprocal is not true: one may find a low value of 
${\cal E}$ in which each player does not belong to any local minority. This 
is because ${\cal E}$ is associated with the balance within the whole ensemble 
of players between those that make one choice and those that make the opposite, 
while $H$ refers instead to the interaction of each player with her neighborhood.     
The spin-spin interaction in Eq.(\ref{eneloc}) is not that of a true antiferromagnet 
because becomes weaker with a larger $n$. This has been chosen in this way to 
resemble more closely the rules of the LEMG, in which  each agent changes accordingly
to the trend in her neighborhood, paying no attention to the actual number of agents 
in each side.
 
A random relaxation dynamics that tends to minimize $H$ consists in updating 
the spins of the system, randomly and asynchronously following the rule:

\beq
s_i(t) \rightarrow s_i(t+1)= -\mbox{sign}[Q_i(t)]
\label{relax}
\eeq  

This dynamics leads to asymptotic configurations that are the same as the 
ones obtained with the LEMG using the random relaxation dynamics in both 1D 
and 2D systems. In the 2D case it is obtained the same pattern of domains 
that accommodate horizontal and vertical stripes of widths that are 
determined by the size $n$ of the neighborhood.

\begin{figure*}[tbp]
\begin{tabular}{ccc}
\includegraphics[width=6cm,clip]{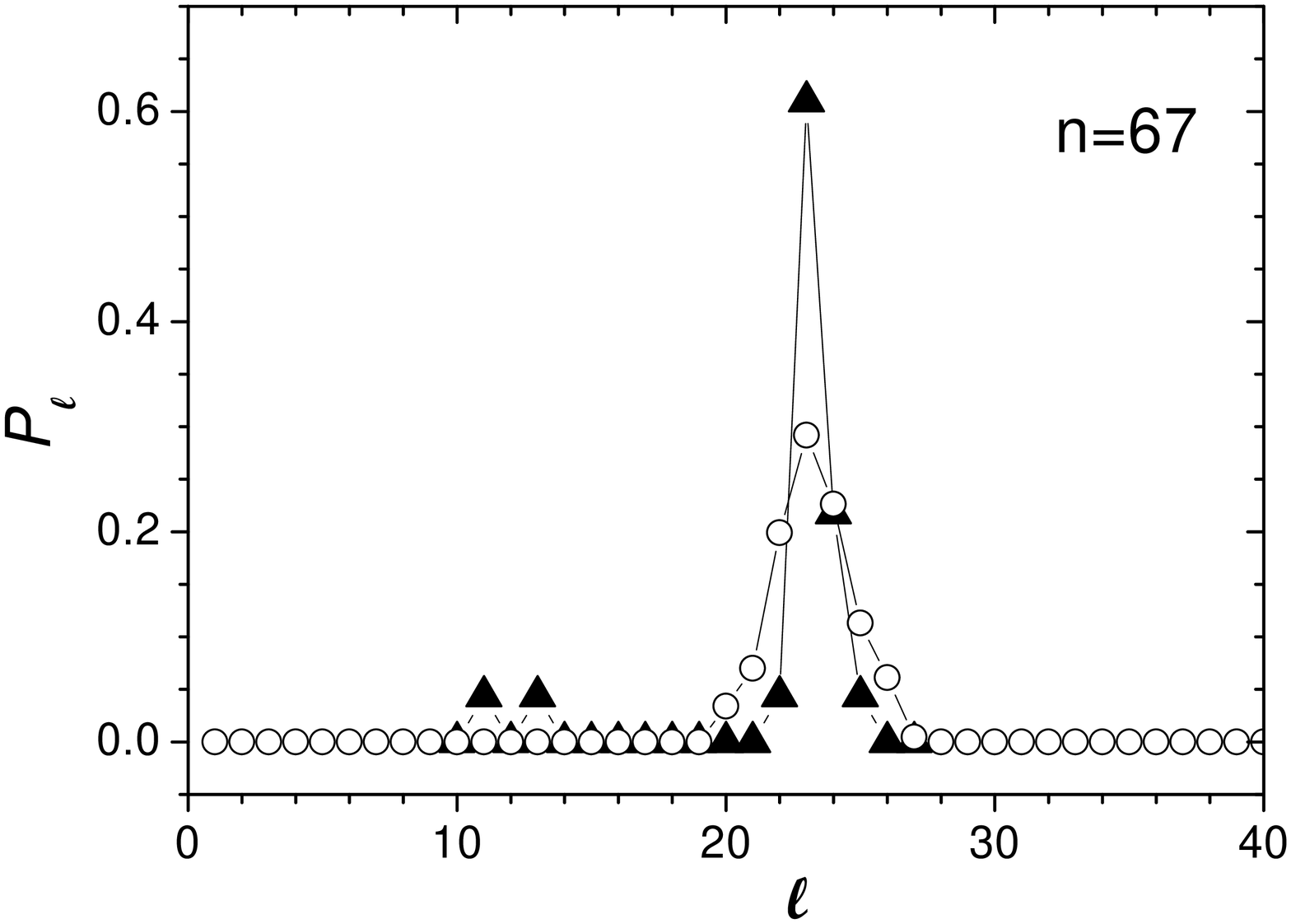}  & \includegraphics[width=6cm,clip]
{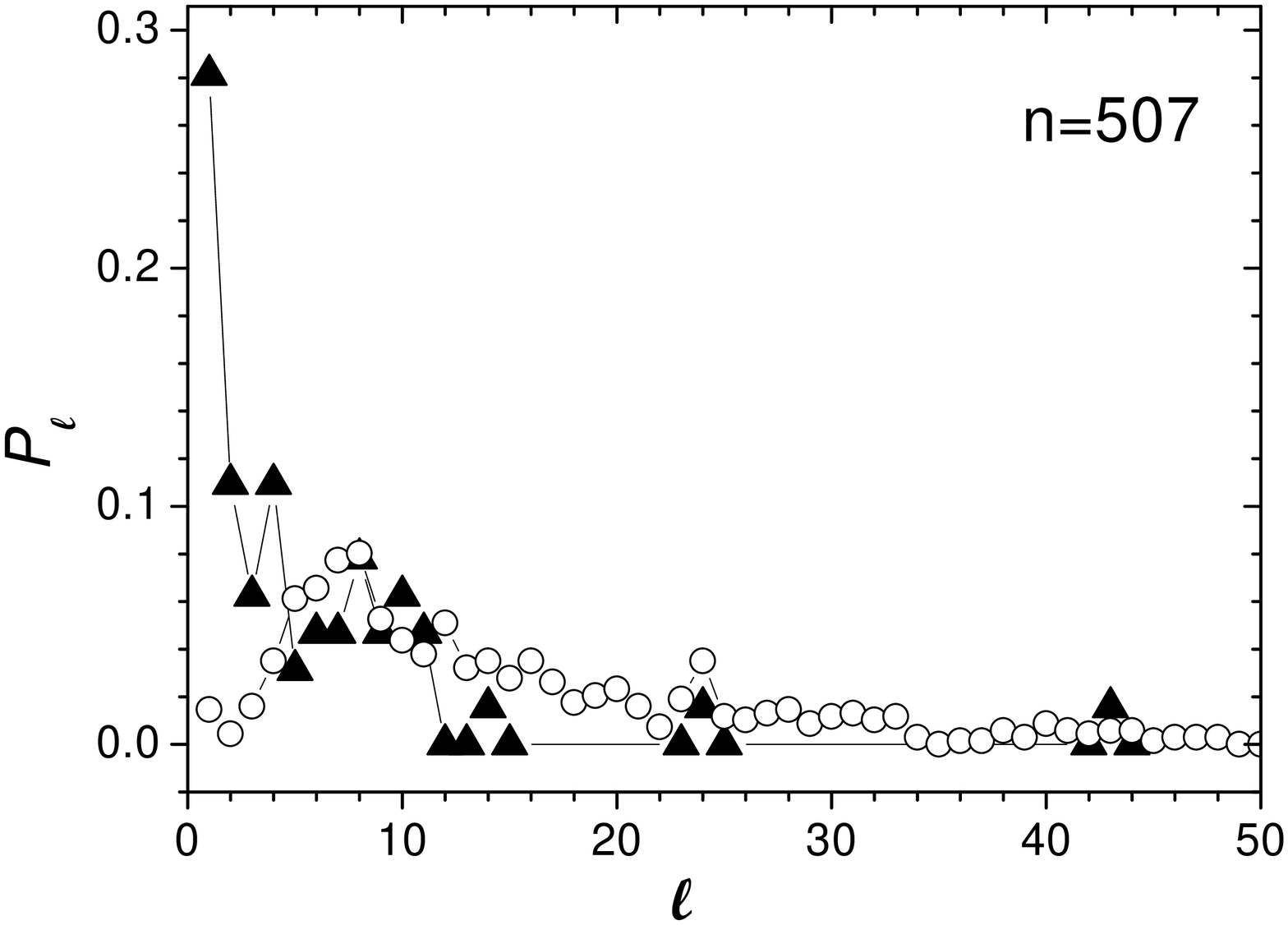} & \includegraphics[width=6cm,clip]{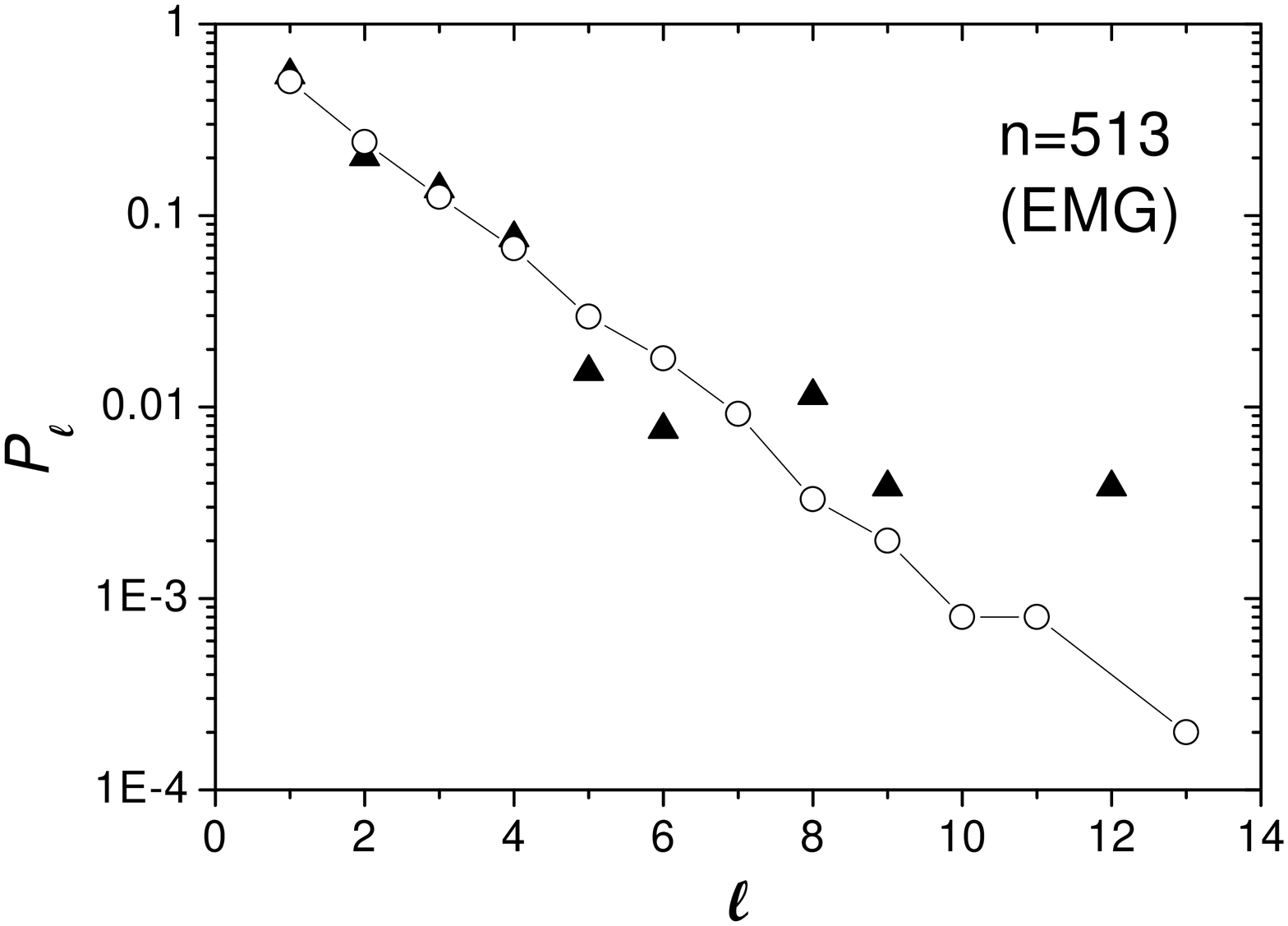} \\
\end{tabular}
\caption{We show the probability distribution of the string length $P_{\ell}$, for
a linear system of 513 agents, for three different neighborhoods. Full triangles 
correspond to the LEMG, while open circles are the results of the spin model. It 
can be seen that the $n$=507 case represents an intermediate situation between that 
of $n$=67 (fully ordered), and that of $n$=513 (fully disordered). Notice that the 
$n$=513 case is plotted in a semilog scale.}
\label{entornos}
\end{figure*}

The distribution of string lengths is displayed by the probability $P_\ell$
of occurrence of a string of length $\ell$. In Fig.\ref{entornos} we 
compare the distribution $P_{\ell}$ obtained for the LEMG minimizing $H$ with
the spin flipping dynamics of Eq.(\ref{relax}), as the size of the neighborhood 
is changed. We show as an example the cases with $N=513$ and $n=$67, 507 and
513. For $n=67$ very few string lengths survive and a fully ordered pattern emerges 
with strings of a length $\ell \simeq n/3 \simeq 23$. 

This noisy spectrum in $P_{\ell}$ for $n$=507 is the signature of some 
degree of disorder of the system. This is, however, a {\em quenched} 
disorder because it appears in the absence of any fluctuation that can be 
attributed to a temperature. In fact as we will immediately see the 
relaxation dynamics of Eq.(\ref{relax}) can be considered to correspond to 
the limit $T \rightarrow 0$ of a thermal relaxation dynamics. The disorder 
should therefore be attributed to the frustration imposed by the boundary 
conditions. In fact it becomes more important when $n/3 > N/4$. \emph{i.e.} 
when the length of the system is no longer able to accommodate an even 
number of alternating strings with a length $n/3$. Such disorder dominates 
when $n=513$ (i.e. $\zeta = 1$) which corresponds to the EMG; in this case 
$P_{\ell}$ merges into a smooth exponential distribution. This agrees with 
the well known correspondence between the EMG with a disordered model of a 
spin glass.

The results displayed above indicate that the LEMG and the spin model 
provide a similar physical insight. In both models, the size of the 
neighborhood is a control parameter that allows to gradually interpolate 
between a fully ordered antiferromagnetic system and a fully disordered, 
spin glass like system. In the following section we study how the 
antiferromagnetic model allows a thermodynamical approach that is difficult 
to envisage as an extension of the EMG.

\section{A thermodynamic treatment}

The occurrence of a pattern of strings of 0's and 1's is a signature of the 
LEMG that is reproduced within the antiferromagnetic model. The system gets 
ordered as a result of the minimization produced by the relaxation 
dynamics. In a general situation also random (thermal) fluctuations should 
be introduced in this process. These correspond to agents that have a 
finite probability of changing their decisions at any time. A low 
temperature can be associated to a very low probability of such change 
while the high temperature limit should be associated to agents that flip a 
coin to make up their minds. If thermal fluctuations are allowed, 
equilibrium corresponds to a minimum of a free energy and this in turn 
involves an energy minimization and an entropy maximization. In the present 
section we concentrate in this picture. We will restrict to consider a 1D 
discrete spin system within the antiferromagnetic model.

\subsection{The string length distribution}

We now discuss how a long range ordering of the system emerges as a 
consequence of the minimization of a free energy. We focus the discussion 
of the present section on the case in which $\lim_{N \rightarrow \infty} 
\zeta = 0$. The effects of a finite values of $\zeta$ as $N\rightarrow 
\infty$ will be discussed in a forthcoming section within the framework of 
the spin model presented above.

Let us assume a sequence of $N$ 0's and 1's randomly selected with equal 
probability. We therefore assume that $N_1 \simeq N_0 \simeq N/2$ up to 
terms $O(1/N)$. In what follows we shall assume that one kind of characters 
(say 0's)  remains frozen and we will find the distribution of the lengths 
of strings of the other type (say 1's). Due to the symmetric situation 
between the 0's and the 1's, what we will conclude for either one is also 
valid for the other. Let $n_m$ be the number of strings with $m$ equal 
characters and let $M$ be the total number of these sequences. Consequently 
the probability of occurrence of a sequence with $m$ characters is 
$P_m=n_m/M$. In addition holds that $\sum_m mn_m= N/2$. We can also define 
an average string length through $\hat{\ell}=\sum_m mP_m$.

In order to construct a free energy we have two tasks, one is to define the 
entropy function and the other is to define the energy. The entropy of a 
configuration of strings can be calculated as the logarithm of the number 
of states in which $M$ strings can be arranged preserving the fact that the 
$n_m$ strings of length $m$ are indistinguishable. Namely

\beq
{\cal S}=\log \Bigl[ \frac{M!}{\Pi_m n_m!} \Bigr]  \simeq -M\sum_m P_m \log (P_m)
\eeq

The energy function $\widetilde{\cal E}$ for the same configuration of 
strings must depend upon the neighborhood structure and therefore must 
carry the information of the kind of \emph{local} game that are 
playing the $N$ agents of the system. If an energy function $E_m$ can be 
defined for a string of $m$ equal characters, the energy of the ensemble of 
strings can naturally be defined as the weighted sum:
\beq
\widetilde{\cal E}=\sum_m P_m \frac{E_m}{\hat{\ell}}
\label{energy}
\eeq
We have introduced $\hat{\ell}$ in the denominator in order to work with 
energies per unit length. We stress the point that $\widetilde{\cal E}$ is 
an energy associated to \emph{an ensemble of strings} and is connected to 
$H$ as given in Eq.(\ref{eneloc}) only by the spin interaction that is used 
to calculate the energies $E_m$.  It is clarifying to consider the 
particular case of $n=3$. In this case the energy $E_m$ can unambiguously 
be evaluated with a spin interaction as the one used in Eq.(\ref{eneloc}). 
Since the sequence of $m$ equal characters is flanked by one character of 
the other kind at both ends, $E_m$ can be calculated as in 
Eq.(\ref{eneloc}) by sliding a window of width $n=3$ along the sequence and 
counting the energy contributed by each position, namely: 
\beq
E_m=\frac{1}{4}\sum_i^m s_m Q_m
\eeq
An alternative possible calculation of $E_m$ is to use a similar sliding 
window and counting in each site if it corresponds to a local winner or 
loser. Notice that this procedure is exact only for $n=3$. For a larger 
neighborhood additional hipoteses must be made concerning the environment 
in which the string is placed. We come back to this point later.

We can now turn to find the distribution of the lengths of the strings. For 
this purpose we need only to evaluate the set $\{ P_m \}$ that correspond 
to a minimum of the free energy per unit length

\beq
F= \widetilde{\cal E} - \frac{T{\cal S}}{N/2}
\label{free}
\eeq

In order to find a minimum of $F$, one still has to add a term with the 
Lagrange multiplier $\lambda ( \sum_mP_m -K)$ to impose the condition that 
$\sum_m P_m =K=$constant. We have introduced the thermodynamic temperature 
$T$ that can be associated as usual to fluctuations in the arrangement of 
strings. We discuss in the next section the microscopic influence of $T$ in 
a spin model of the LEMG.  

In order to calculate the minimum of $F$ we set:
\beq
\frac{\partial F}{\partial P_{\ell}}=
\frac{\partial \widetilde{\cal E}}{\partial P_{\ell}} + \frac{T}{\hat{\ell}^2} 
\Bigl[ \hat{\ell}(1+\log P_{\ell})-\ell \sum_m P_m\log P_m \Bigr]+\lambda = 0
\label{derivada}
\eeq
  
This equation can be solved for $\lambda$. In order to do so we first 
eliminate the explicit dependence on $P_{\ell}$ by multiplying by 
$P_{\ell}$ and performing a summation over $\ell$. In this operation one 
can impose that $\sum P_{\ell} = K$ but, as expected, the normalization 
constant $K$ remains undetermined. Solving for $\lambda$ we obtain:

\beq
\lambda = -\frac{T}{\hat{\ell}} - \langle \partial \widetilde{\cal E} \rangle 
\label{lambda}
\eeq
with $\langle \partial \widetilde{\cal E} \rangle = \sum_m P_m \partial 
\widetilde{\cal E}/\partial P_m$. Next $\lambda$ can be replaced in Eq.(\ref{derivada}) 
to solve for $\log P_{\ell}$. We finally obtain:
\beq
P_{\ell} = \frac{1}{K}\exp [-\ell {\cal S}/\hat{\ell}M ] \exp[-\hat{\ell}\Delta_{\ell} /T]
\label{pq}
\eeq
where $\Delta_{\ell}= \partial \widetilde{\cal E} / \partial P_{\ell} - \langle \partial 
\widetilde{\cal E} \rangle $. Te normalization constant $K$ is determined with the 
condition that $\sum P_{\ell} =1$ because the $P_{\ell}$'s are probabilities of mutually 
excluding events.

The Eq.(\ref{pq}) provides the desired distribution of string lengths. It is 
interesting to consider the limits $T \rightarrow \infty$ and $T 
\rightarrow 0$. The high temperature limit yields a distribution that dies 
exponentially with the length $\ell$ of the strings as measured in units of 
the average length $\hat{\ell}$.  This is indeed the distribution found for 
the (random) distribution used as an initial state for the relaxation 
process used in the LEMG. As we shall see in the next subsection, this is 
also found in the high temperature limit of a multi-spin model of the LEMG. 

An exponential tail in the distribution is the signature of some kind of 
disorder. This can either be due to thermal fluctuations as in the present 
considerations or correspond to a quenched disorder of the system. The latter
has been reported in the previous section as the result of the relaxation process 
when the size of the neighborhood becomes equal to the size of the whole system. Such 
quenched disorder of the EMG also displays a distribution that has an exponential 
distribution of string lengths. 

To investigate the limit of $T\rightarrow 0$ we use Eq.(\ref{energy}) and find that 
\beq
-\frac{\hat{\ell}\Delta_{\ell}}{T} = -\frac{E_{\ell}-\widetilde{\cal E}}{T}=
-\sum_mP_m \frac{(E_{\ell}-E_m)}{T}
\eeq

This expression allows to find immediately the asymptotic distribution of 
lengths in the limit $ T \rightarrow 0$. Assume then that the set of the 
$\{ E_m \}$ is ordered $E_{m_1}<E_{m_2}<\cdots <E_{m_N}$. One can easily 
recognize that for $T \rightarrow 0$, and $\ell= m_N$ the probability 
$P_{\ell}\rightarrow 0$ because all exponents $-(E_{\ell}-E_k)/T$ with 
$k=m_{N-1},m_{N-2}, \cdots ,m_1$ in the second exponential of Eq.(\ref{pq}) are 
negative causing $P_{\ell}$ to approach 0 exponentially for low $T$. A 
completely similar argument holds for $\ell= m_{N-1}, m_{N-2}, \cdots m_2 $. The only 
term that survives is $\ell= m_1$ the one corresponding to lowest energy 
$E_{m_1}$ that, due to the normalization imposed on the probabilities, must 
fulfill $P_{m_1} \rightarrow 1$. We thus see that in the low temperature 
limit the system gets organized by repeating only strings of a given length 
$\ell = m_1$ that corresponds to minimum of $E_{m_1}$. This is what has 
been obtained both in the simulations reported  and within the spin model. 
As we have seen before the length $\ell =1$ corresponds to $n=3$. In 
general the surviving strings with a minimal $E_s$ are of a length that is 
close to one third the size of the neighborhood.

Note that an energy $E_m$ can only be defined without ambiguity for the 
case of $n=3$. This is so because the local energy of the first and last 
character of the sequence is completely defined. If the environment is 
larger, in order to calculate $E_m$ further considerations have to be made 
about the environment in which the sequence is placed. For larger 
neighborhoods the energy of longer strings can't be defined. In such a case 
an average over all possible environments could be made. However, such 
averaging procedure loses meaning as $\zeta $ approaches 1 and the 
considerations made for the low temperature limit are no longer applicable.

\begin{figure}[tbp]
\begin{tabular}{c}
\includegraphics[width=8cm,clip]{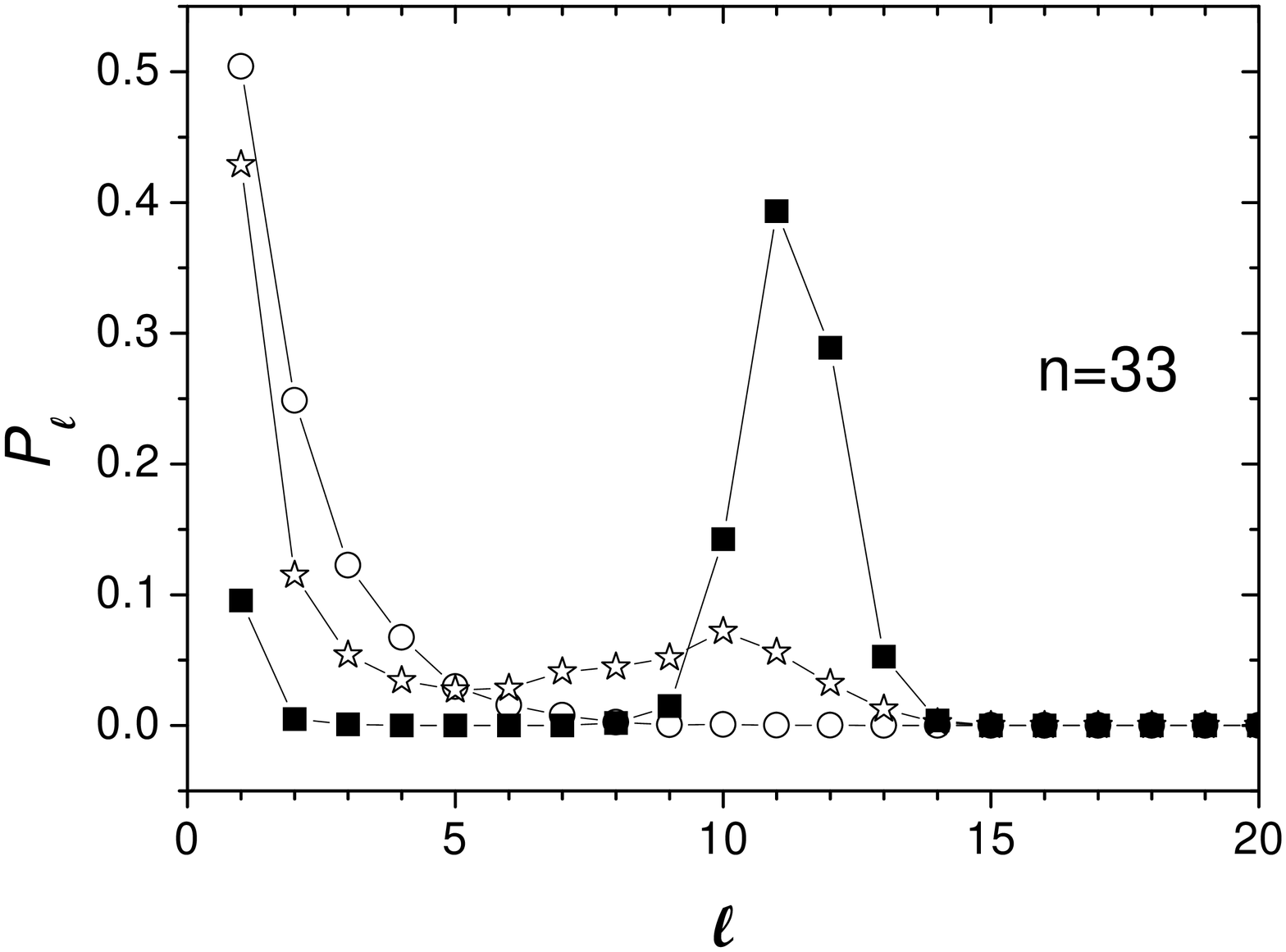} \cr \includegraphics[width=8cm,clip]{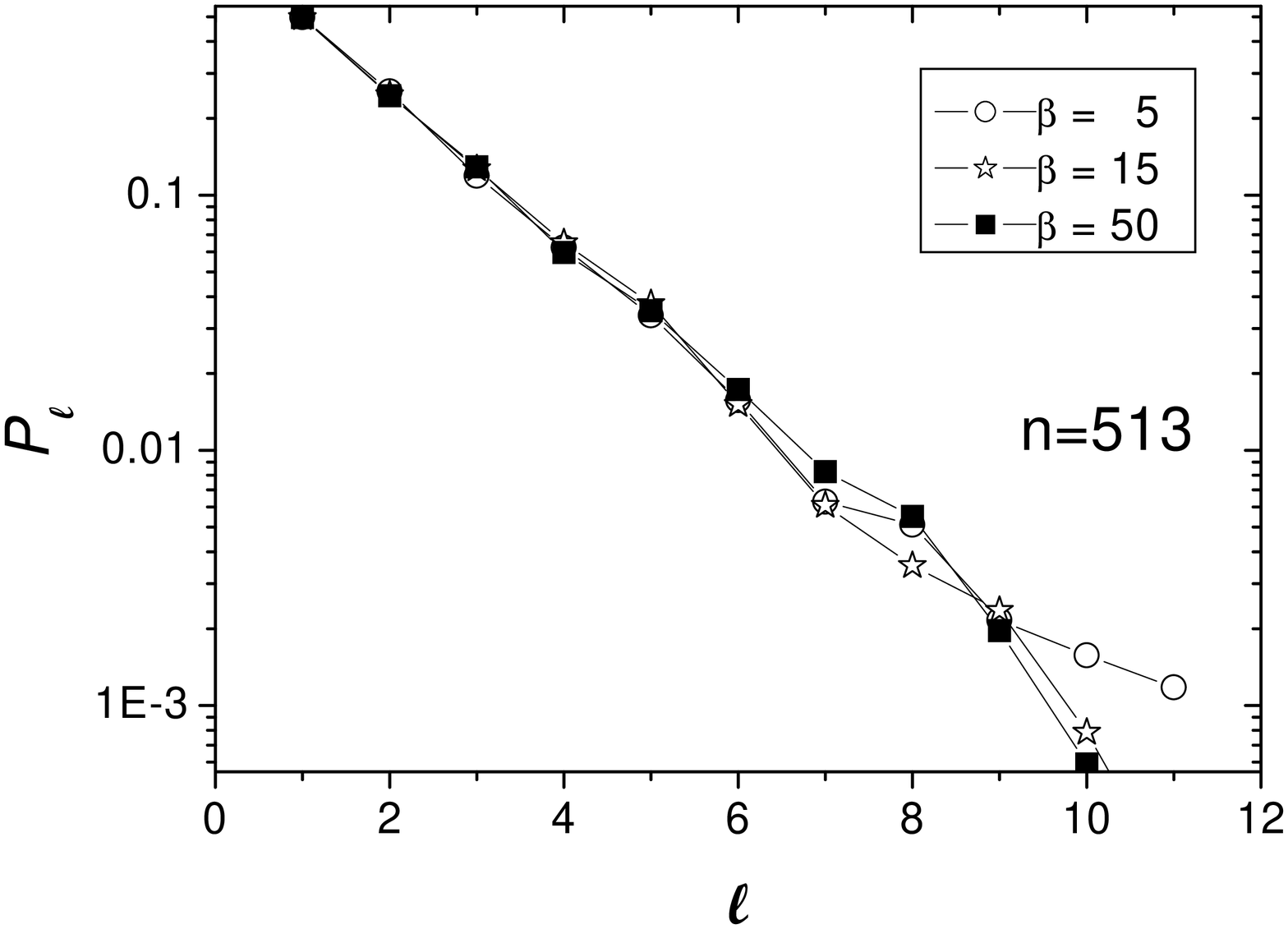} \\
\end{tabular}
\caption{We show the probability distribution of the string length $P_{\ell}$,
for a linear system of 513 agents, for $n$=33 and $n$=513 (EMG), and three 
different temperatures $T=1/\beta$.}
\label{termentornos}
\end{figure}

\subsection {The effects of a finite $\zeta$ }
\label{finito}

\subsection {Thermal fluctuations in the antiferromagnetic model}
 
A microscopic interpretation of the temperature parameter introduced in the
preceding section is desirable. One possibility is to follow the line suggested 
in Ref.\cite{Thermal}. It has been shown there that the random relaxation used 
in the LEMG gives rise to fluctuations that can be assimilated to the introduction 
of a temperature parameter. The fact that the individual probabilities $\{ p_i \}$ 
are updated by choosing the new value at random from an interval $[p_i-\delta p,
p_i+\delta p]$ with reflective boundary conditions at $p_i = 1$ and $p_i = 0$ 
produces fluctuations in the energy function ${\cal E}$ given in Eq.(\ref{costo2}). 

That way is however not acceptable. The introduction of a finite temperature into the 
LEMG amounts to allow for ``hesitating agents'' that may change their current 
decision with a finite probability. The well known gradual updating rule of individual 
probabilities is unable to handle this situation. Within the antiferromagnetic 
model for the LEMG, such changes can easily be assimilated to a probabilistic flipping 
of spins in Eq.(\ref{relax}). This temperature 
parameter is therefore different from the one introduced in \cite{Thermal} because 
both are associated to a different kind of fluctuations. 

To deal with a finite temperature we proceed as usual, namely by assuming 
that a spin may flip with a probability that depends upon $\beta = 1/T$ through:
\beq
\mbox{Prob}(s_i \rightarrow -s_i)=\frac{1}{1+ e^{-\beta s_iQ_i}}
\label{termrelax}
\eeq 
that contains Eq.(\ref{relax}) in the limit of $\beta \rightarrow \infty$ ($T\rightarrow 0$).
 
The ordered pattern for the 1D system in which few string lengths are 
strongly preferred only holds in the low temperature limit. The 
fluctuations introduced through the probabilistic flipping of 
Eq.(\ref{termrelax}) give rise to a thermal disorder that shows itself 
through the distribution function $P_{\ell}$. This competes with the 
quenched disorder produced for large values of the size parameter. As 
expected, in the high temperature limit $P_{\ell} \simeq \exp (-\alpha 
\ell)$.  

In Fig.\ref{termentornos}a we show how the ordered pattern of $n=33$ 
melts down as $\beta \rightarrow 0$ while for $n=513$ 
(Fig.\ref{termentornos}b) the exponential distribution does not change. The 
peak in $P_{\ell}$ for $n=33$ that represents an ordered pattern, gradually 
disappears, being replaced by an exponential for high $T$. The quenched 
disorder that appears for $n=513$ is independent of $T$.

The local energy function given in Eq.(\ref{eneloc}) is sensitive to the 
emergence of a local ordering. The relaxation indicated in 
Eq.(\ref{termrelax}) allows to perform a minimization of $H$ while keeping 
control of the temperature. This allows to investigate the persistence of 
the local ordering and the competition between quenched and thermal 
disorder  as the size of the neighborhood $n$ approaches the size $N$ of 
the whole system.

\begin{figure}[tbp]
\includegraphics[width=9cm,clip]{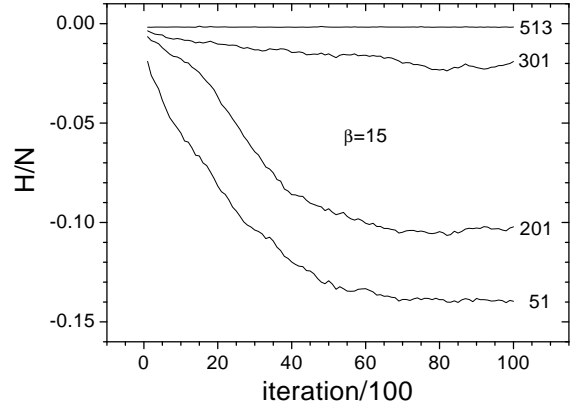} 
\caption{Evolution of the local energy per site, from the minimization process
on the spin model. The number of agents is $N$=513; The values of $n$ are show in the 
figure}
\label{terminimo}
\end{figure}

In Fig.\ref{terminimo} we show the result of the minimization process for 
two different values of $\zeta$. As long as $\zeta < 1$ the local energy is 
lowered as the relaxation proceeds, showing that a local ordering emerges. 
This reduction gradually becomes less important as $\zeta \rightarrow 1$ 
and vanishes completely in that limit. This puts in evidence that when 
$\zeta =1$ the relaxation process is unable to introduce any ordering into 
the system.

\begin{figure}[tbp]
\includegraphics[width=9cm,clip]{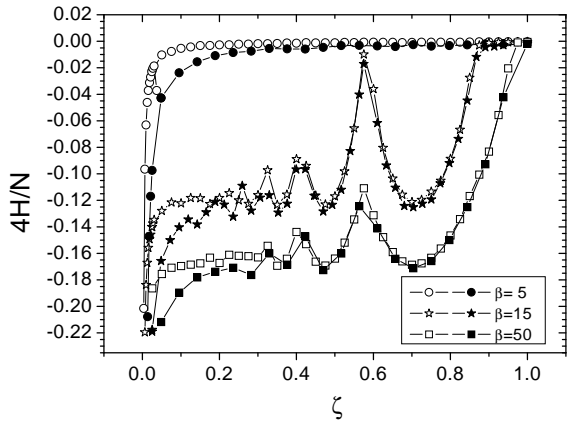}
\caption{Energy per site as a function of $\zeta$,for three different values 
of $T=1/\beta$. Empty (full) symbols correspond to $N$=2001 (513) agents.}
\label{enetorno}
\end{figure}

The energy function given in Eq.(\ref{eneloc}) is expected to show some 
degree of structure as the size of the neighborhood changes. This is due to 
the frustration caused by the competing boundary conditions that arise from 
the concatenation of local minorities in neighborhoods that are comparable 
to the size of whole chain. Such structure can clearly be seen in the plot 
of $H/N$ vs $\zeta$ that is shown in Fig.\ref{enetorno}. This pair of 
variables have been chosen because the corresponding curves turn out to be 
independent of $N$. The structure that is displayed must therefore be 
considered to be independent of the size of the system. 

String lengths that are even fractions of $N$ minimize the frustration 
because they better fulfill periodic boundary conditions of the whole 
system by accommodating an even number of strings of equal length along the 
chain. One therefore  expect that the average string length must be related 
to $N$ through $\hat{\ell}=N/4, N/6, N/8$ etc. On the other hand for an 
alternating pattern of +1's and -1's, the average string length must in 
turn related to the size of the neighborhood through $\hat{\ell}=n/3$, in 
order to achieve a minimal energy. An optimal ordering of the system is 
therefore possible for $\zeta \simeq 3/4, 3/6, 3/8,$ etc. These values are 
seen to correspond to minima of the curves in Fig.\ref{enetorno}. The 
maxima in between correspond to a missmatch between the value of the  size 
of the neighborhood and the size of the whole system, causing that such 
optimization can not be achieved. This structure is seen not to survive to 
thermal fluctuations. 

\section{Conclusions}

The MG has been greatly used as a working model for self-organization. 
Individual players, without any central coordination are nevertheless able 
to minimize a cost function that can be assimilated to an energy, by self-
segregating into two different groups that make opposite elections. This 
features is neatly displayed by the density distribution of individual 
strategies that shapes as a U. 

If the rules of the game are extended by placing the agents into the nodes 
of a regular or a random graph several important changes occur. In the 
first place the density distribution of individual strategies turns out to 
be insensitive to the neighborhood structure of the new model. In the 
second place a long range spatial ordering dominates the system. In the 
third place self-organization is improved by producing lower value of the 
energy function. 

These features can fruitfully be studied in 1D systems. We found that  the 
distribution of strings of players that make similar options is a relevant 
parameter that provides a signature of the type and kind of order that is 
induced in the system. This distribution changes between a delta-like 
function in which strings of a few lengths are allowed to survive to an 
exponential distribution of string lengths that is a signature of 
randomness. This transition puts in evidence that the LEMG provides a 
gradual interpolation between an ordered system in the limit of small 
neighborhoods and a fully disordered system when the size of the 
neighborhoods approaches de size of the system. 

We have shown that the LEMG is equivalent to a  spin system with an 
antiferromagnetic-like interaction that extends to a neighborhood whose 
size is taken as a control parameter. This model displays the same pattern 
of local ordering and string length distribution and also displays the same 
order-disorder transition when the range of the antiferromagnetic 
interaction approaches the size of the whole system. In addition an energy 
function can be introduced that can be considered as an extension of the 
LEMG in which the neighborhood structure is explicitly accounted for. A 
minimum of such energy leads to a minimum of the energy function used for 
the LEMG but the converse is not true.

The distribution of string lengths can also be studied in the thermodynamic 
limit of an infinite linear system of spins in a thermal bath. The 
temperature parameter that has to be introduced is absent in the LEMG and 
corresponds to hesitating agents that may change their decision with a 
finite probability at any time. Using this model we were able to find the 
high and low temperature limits of the probability distribution of string 
lengths as arising from a minimization of a free energy. The low 
temperature limit, merges for small neighborhoods into the one derived from 
the usual relaxation dynamics of the LEMG or the antiferromagnetic model 
yielding a delta-like distribution of lengths in which only very few 
privileged lengths survive and a long range ordered pattern emerges. 

The high temperature limit can only be treated within the antiferromagnetic 
model. Within this framework an order-disorder transition can be predicted. 
Thermal randomness associated to the high temperature limit has the same 
exponential distribution of string lengths that is numerically obtained in 
the quenched disordered limit for zero temperature. We have also studied 
the effect of the interplay of the size of the system and the size of the 
neighborhood. We have shown that the energy of the spin system as a 
function of the size parameter can be cast into a form that is independent 
of the size of the system. This function that is highly structured puts in 
evidence the occurrence of ``preferred'' and ``hampered'' values of the 
size parameter that correspond to minima and maxima of the frustration 
introduced by conflicting boundary conditions.

The 1D system that we have considered in the present paper allows to get a 
detailed physical picture of the LEMG. It helps to understand how a 
structure of neighborhoods makes it possible a better ordering of the 
system that is in turn reflected as a minimum of the energy. Such ordered 
phase that is  equivalent to an antiferromagnetic ordering shows up as a 
better coordination of the decisions made by decentralized agents in a 
minority game.


\begin{thebibliography}{ }
\bibitem{foot} This distribution is actually the composition of two asymmetric 
distributions, one that vanishes for $p<1/2$ and the other for $p<1/2$.
\bibitem{foot2} This correspondence is exact only in the limit of a perfect 
self-segregation.
\end{thebibliography}
\end{document}